**Light induced resistive switching in copper oxide thin films**

L.D. Varma Sangani[1,2], M. Ghanashyam Krishna[1,*]

[1]Centre for Advanced Studies in Electronics Science and Technology, School of Physics, University of Hyderabad, Hyderabad-500046, Telangana, India
[2]Tata Institute of Fundamental Research, Mumbai, India
[*]Corresponding author: email address: mgksp@uohyd.ernet.in

**Abstract**

Copper oxide based metal-insulator-metal structures were subjected to white light irradiation. The top electrodes included Al, Cr and Ni while the bottom electrode was either Au or Pt. A white light pulse controls the set process and this light – induced set (LIS) can be performed at very low voltage (tens of milli volts) which is not possible in the normal set process. The LIS is initiated at the positive edge of the pulse, and there is no effect of falling edge of the light. In most cases the high resistance state (HRS) to low resistance state (LRS) transition is irreversible i.e. the devices continue to remain in the LRS even after the light pulse is switched off. Light induced reset is achieved in only one device structure, Al/Cu$_x$O/Au. By using LIS and LIR, set and reset power of the device can be reduced to a great extent and the set and reset parameters variation also reduces. The current work, thus, points to the possibility of formation and compliance-free resistive random access memory devices.

**Introduction**

Change in conductivity by means of photon irradiation on a semiconductor is a well known phenomenon. Many devices such as photovoltaic cell (PVC), light dependent resistor (LDR) and light sensors work on this principle which is attributed to electron – hole pair generation due to irradiation[1]. The electrons present at the top of the valence band absorb the incident photon energy and are excited to the conduction band creating electron – hole pairs leading to increase in conductance of the semiconductor[1]. The study of light induced resistive switching as an approach to control the memory states in a RRAM has been studied by few groups. Switching parameters of RRAM were observed under white and UV light [2-10]. For example, some authors reported increase in LRS and HRS resistances while others reported decrease in both resistance values [2,3,6,7]. There are, however, very few reports on controlling the switching parameters such as $V_{set}$, $V_{reset}$ and $I_{reset}$.

In this paper, the phenomenon of light induced resistive switching in $Cu_xO$ based RRAM devices is investigated. It is demonstrated that control over the switching parameters of the device can be achieved by pulsing the white–light beam.

**Experimental setup**

Vertically stacked metal (Top) /oxide/metal (Bottom) structures of 20x20 μm$^2$ area were fabricated on Silicon substrate; D1 with Al/$Cu_xO$/Au[3], D2 with Cr/$Cu_xO$/Au, D3 with Ni/$Cu_xO$/Au and D4 with Ni/$Cu_xO$/Pt. The Au film, of 100 nm thickness, is deposited by thermal evaporation at a pressure of $5.0 \times 10^{-5}$ mbar. The Cu film, of 50 nm thickness, is deposited by RF magnetron sputtering from a cathode of 50 mm diameter in a pure Ar atmosphere at a pressure of $1.8 \times 10^{-2}$ mbar followed by post-deposition annealing in air at 250°C for 4 hours to obtain the oxide film. The top electrodes, Cr and Ni of 80 nm thickness were then deposited on the thermally oxidized Cu layer by e-beam evaporation at a pressure of $1.4 \times 10^{-6}$ mbar and $8.6 \times 10^{-6}$ mbar respectively. The Pt bottom electrode for D4 is deposited using RF magnetron sputtering with the same condition used for Cu film. Device D1 fabrication conditions are mentioned in the previous work[11]. Photo lithography technique is used to fabricate all the devices.

The lithography process flow is resist patterning, deposition and lift-off. A positive tone i-line photo-resist (Fujifilm OIR 620 – 10M) was used and an ultraviolet (UV) mask aligner (MJB4 of Suss Microtech) was used for photoresist exposure. The current–voltage (I–V) characteristics, endurance and retention measurements of the RRAM cells were carried out using a semiconductor device analyser (Agilent B1500A). A voltage sweep was used for current–voltage measurements as well as for the endurance property measurements of the memory cell. The sweep was applied to the top electrode and the bottom electrode was grounded. A voltage pulse of 100mV with 100μs width was applied for every 60s to observe the retention measurement.

The setup used for studying the light – induced Resistive switching is shown in fig. 1. A white ring fluorescent light source (12W, 220V) and an LED source (~12W, 12V) are attached to the probe station(S1160) to record the I-V characteristics under illumination. The toggle switch (S) attached to the light source is used generate the pulse, by switching on and off the source manually with a fixed on time of ~1s for all experiments in this work.

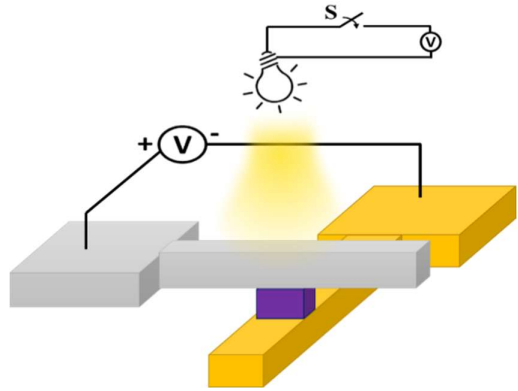

Fig 1 Setup for Light – induced resistive switching

**Results and Discussion**

The results of light – induced measurements carried out on the Chromium based $Cu_xO$ device of structure $Cr/Cu_xO/Au$ are shown in fig. 2 (a)-(d). The set and reset curves without illumination are displayed in fig 2(a), (b) while the response to illumination is shown in fig. 2 (c), (d). The response to illumination is observed by switching on the light pulse during the I-V measurement cycle for 1sec and then switching it off. An example of such a measurement is displayed in fig. 2(c). It is evident that there is a transition to the LRS when the light pulse is switched on. Significantly, the transition from HRS to LRS is irreversible, i.e. the device stayed in the LRS even after the light is switched off. This light induced set (LIS) was repeated during different cycles of I-V measurements and it is evident from the inset in fig. 2(c) that the process is reproducible for successive cycles. The device can be reverted to the HRS only by taking it through the reset process as shown in fig. 2(d). In contrast to the set process, the reset process is not affected by the presence of the light pulse. This is also clear from fig. 2(d) wherein several light pulses are shown during the reset process.

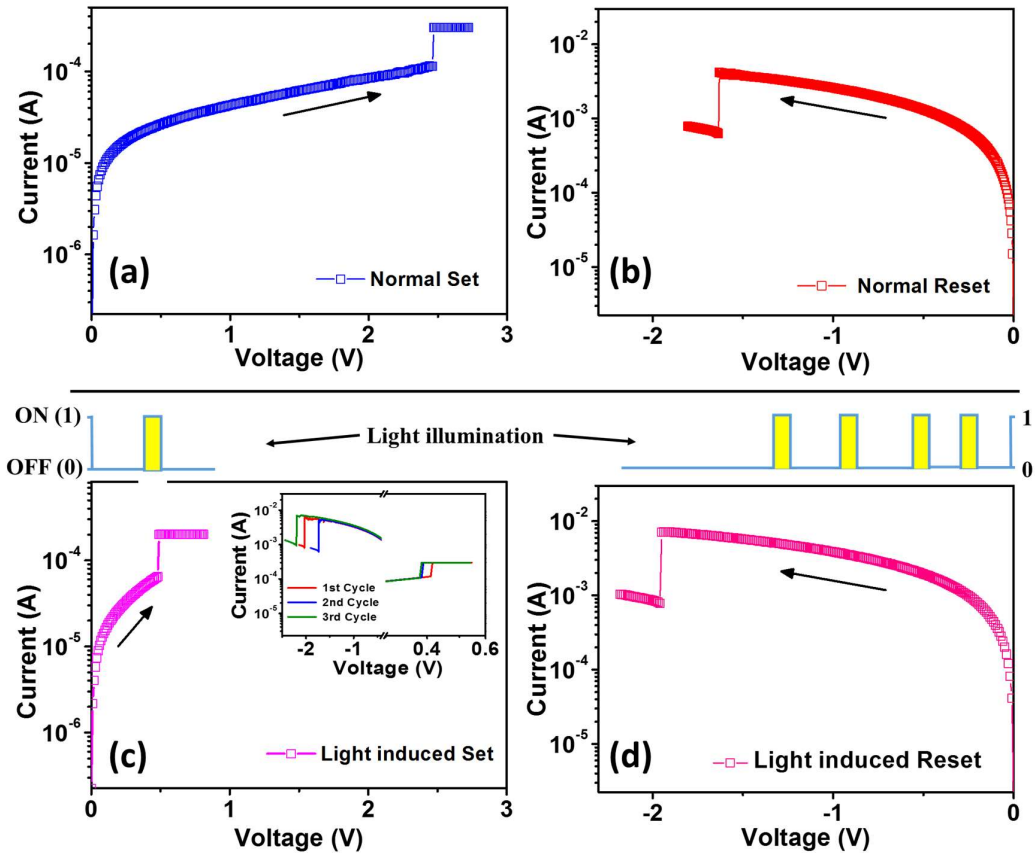

Figure.2: I-V measurement of set and reset of Cr/Cu$_x$O/Au (a), (b) without light illumination and (c),(d) with light illumination, respectively.

It is to be noted that, in fig. 2(c), the light pulse is introduced at ~0.4V which is much lower than the 2.5V for the set process in the absence of light. To determine if the LIS could be achieved at a lower voltage, the light pulse was switched on when bias in the I-V measurement was ~50 mV. Indeed, even at this low voltage as shown in fig. 3(a), there is LIS. There are reports where the set and reset of the device reduced under light illumination [12]. In the present case, however, the set voltage of the device is fully controlled by the light pulse. More significantly, the light pulse not only reduced the variation in the set voltage of the device but also the set process power due to its low set voltage and low compliance current. As in the previous case reset is achieved by going through the normal reset process, as observed from fig. 3(b). Thus, set can be achieved by applying a light pulse at any voltage and reset by the normal reset process. More significantly, LIS could be achieved at any pre-determined voltage.

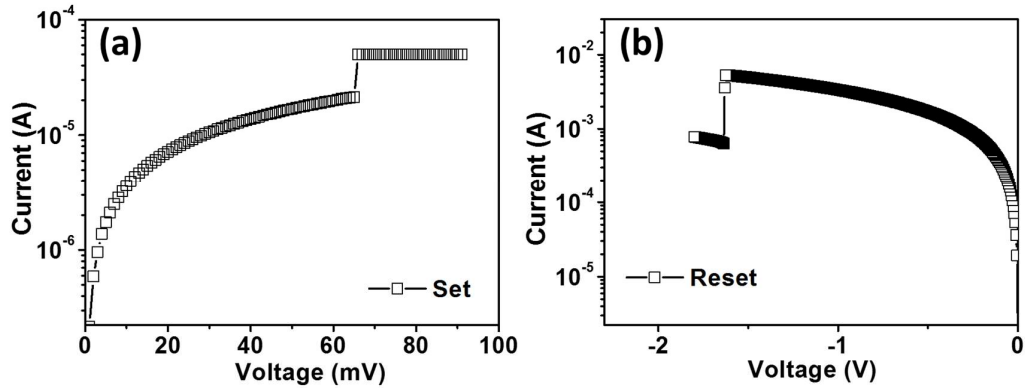

Figure 3 (a) Low voltage light–induced set and (b) immediate reset of Cr/Cu$_x$O/Au.

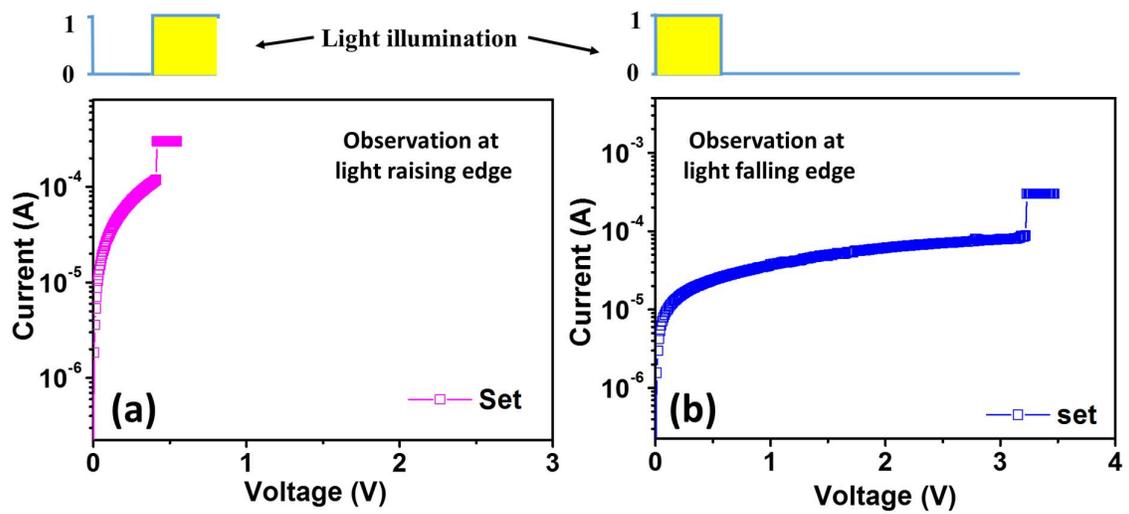

Figure 4 Set observation at (a) rising edge and (b) falling edge of the light pulse.

One of the questions to be addressed is whether the LIS is occurring at the rising edge of the pulse or the falling edge. To determine this, the set process was observed in two more conditions. In the first condition light is switched on while the set sweep is running and in the second condition, light is switched on before set sweep is started and switched off while the sweep is running. The set process under these conditions is displayed in fig.4(a) and (b), respectively. It is evident from fig. 4 (a) that the rising edge of the light pulse controls the set process.

An additional advantage of light induced resistive switching is that it makes the device compliance free. This is a consequence of the very low LIS voltage, as shown in fig.5(a), thus avoiding device damage and also extra circuit for current compliance. The complete set and reset cycle with a single voltage sweep from 1V to -2V is depicted in fig. 5(b). In this experiment the voltage sweep is started at 1V and increased to 2 V in steps of 10 mV and reversed back to 1V. In the figure, cycles 1 to 4 represent the forward sweep, and cycles 5 to 8, the reverse sweep. Initially, the device is in the HR state (Cycle 1) and a light pulse is applied at 0.7V which results in a set for the device and transition into the LR state. Thereafter, the device followed the low resistance path (Cycle 2 and 3) and switched back to HR state at -1.1V and remained in the HR state on further voltage cycling (Cycle 4 to 8).

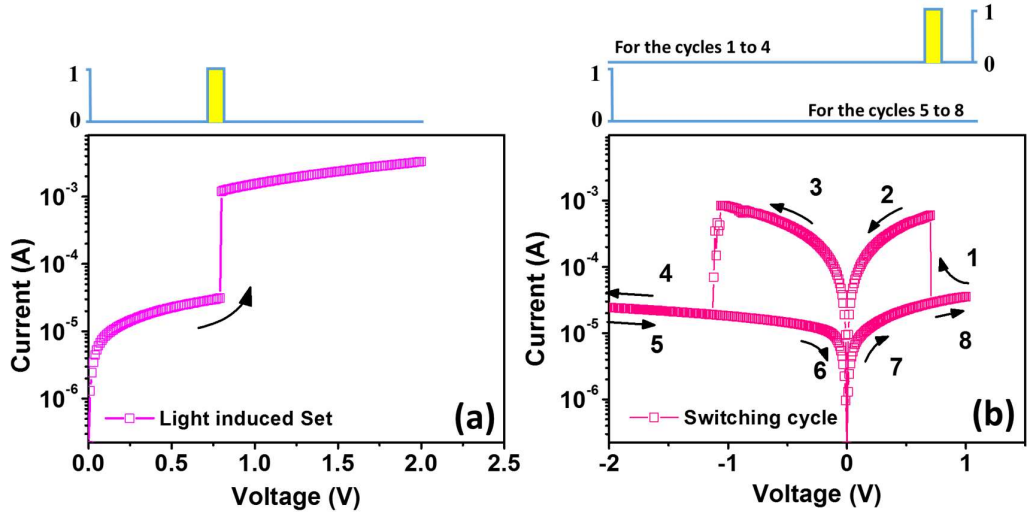

Figure 5 (a) Set process without compliance current, (b) light – induced Switching cycles hysteresis

This property of irreversibility is very attractive for data protection applications [13,14]. The effect of varying electrode materials, on light induced resistive switching was also investigated. The device structures studied were Au/Cu$_x$O/Au, Al/Cu$_x$O/Au, Ni/Cu$_x$O/Au and Ni/Cu$_x$O/Pt. It is evident from fig.6(a)-(d) that all the device structures exhibit light induced resistive switching. As in the case of the Cr based devices described earlier, the LIS occurred at a much lower voltage (~1 V) than the normal set voltage (~2.5V). Similar to Cr based devices, these devices also did not respond to the light pulse in reset process.

The lowest LIS voltage observed in Al, Au, Cr and Ni top electrode devices are 340, 100, 65 and 41mV respectively. Evidently, the Ni electrode based device showed lowest LIS and Al based device showed highest LIS.

It is clear from the results presented above that, after the device makes a transition from HRS to LRS under the influence of the light pulse, it does not revert back to HRS on switching off the light pulse. The device can be reset only by a normal reset process, i.e. by applying the appropriate bias. However, in the Al based device reset was also controlled by light as shown in fig.7 (a)-(c). LIS and LIR were observed at defined voltages, indicates that reset was achieved at very a low voltage -16 mV. The retention measurement performed for more than $10^4$ s for both the states is displayed in fig.7 (d).

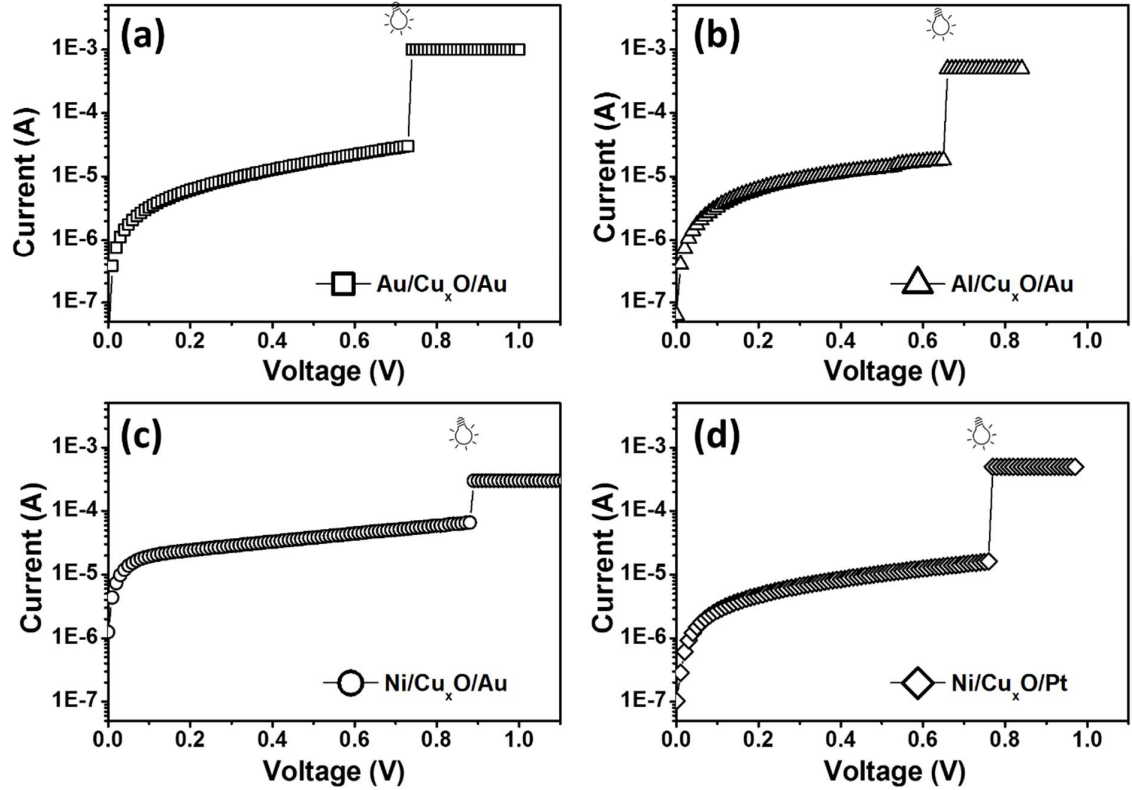

Figure 6 LIS of a)Au/Cu$_x$O/Au, b)Al/Cu$_x$O/Au, c) Ni/Cu$_x$O/Au and d) Ni/Cu$_x$O/Pt.

The results presented in the previous sections lead to the following inferences:

1. Since the LIS is metal electrode independent, the effect is occurring within the semiconductor/dielectric i.e. Cu$_x$O.
2. The magnitude of the change in resistance is electrode dependent.
3. The HRS to LRS transition occurs at much lower voltages than the normal set process and it is irreversible in most cases.

An additional observation was that in many cases the LIS process did not occur during the first pulse. Typical results are presented in 8(a)-(d) and the overall results are summarized in table 1. It is seen that the HRS-to-LRS transition in the Au/Cu$_x$O/Au case occurred in the 5$^{th}$ pulse while in the case of the Al/Cu$_x$O/Au and Cr/Cu$_x$O/Au the transition took place at the 3$^{rd}$ pulse. The behavior of the Ni/Cu$_x$O/Au structure was similar to that of the

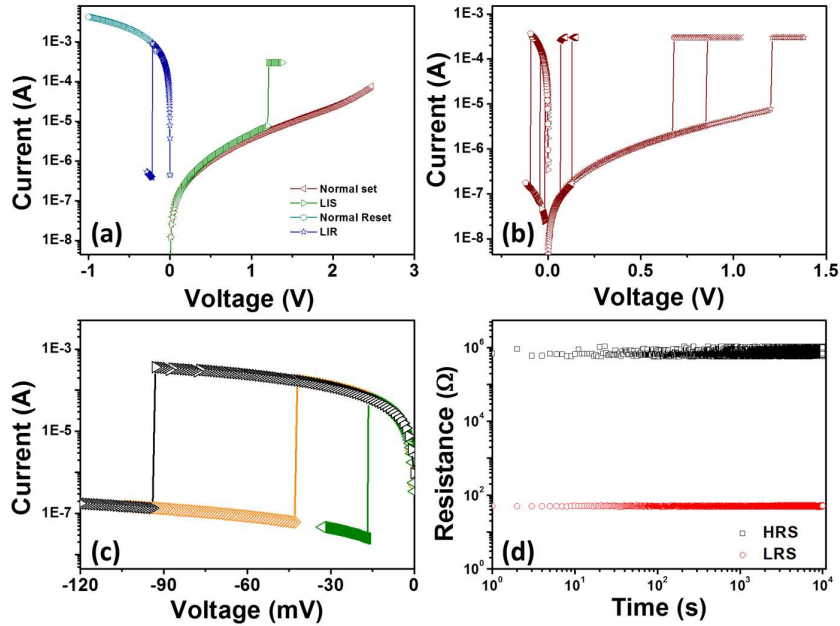

Figure 7 (a) LIS and LIR, (b) LIS and LIR at different voltages, of Al/Cu$_x$O/Au device, (c) zoomed LIR portion of b, and (d) Retention of two resistance states R$_{HRS}$ and R$_{LRS}$ dues to LIS and LIR.

Au/CuxO/Au, in that the HRS-to-LRS transition occurred at the 5$^{th}$ pulse. Interestingly, during the intermediate pulses, i.e. pulses below the threshold number, there is transition back to the HRS state on switching off the light pulse. However, there were also a number of devices that underwent LIS in the first pulse itself. This behavior is quantified in table 1 wherein the total number of times (termed as LIS cycles) the light pulse is incident on a given device is presented. The number of times the device switched from HRS to LRS in the first pulse, termed as N$_{LIS(1)}$, and the percentage of such devices, termed as N$_{LIS(1)}$% are also given. Au and Cr based devices showed more than 80% N$_{LIS(1)}$. Even though the Ni based device showed low LIS voltage, the N$_{LIS(1)}$% is very less.

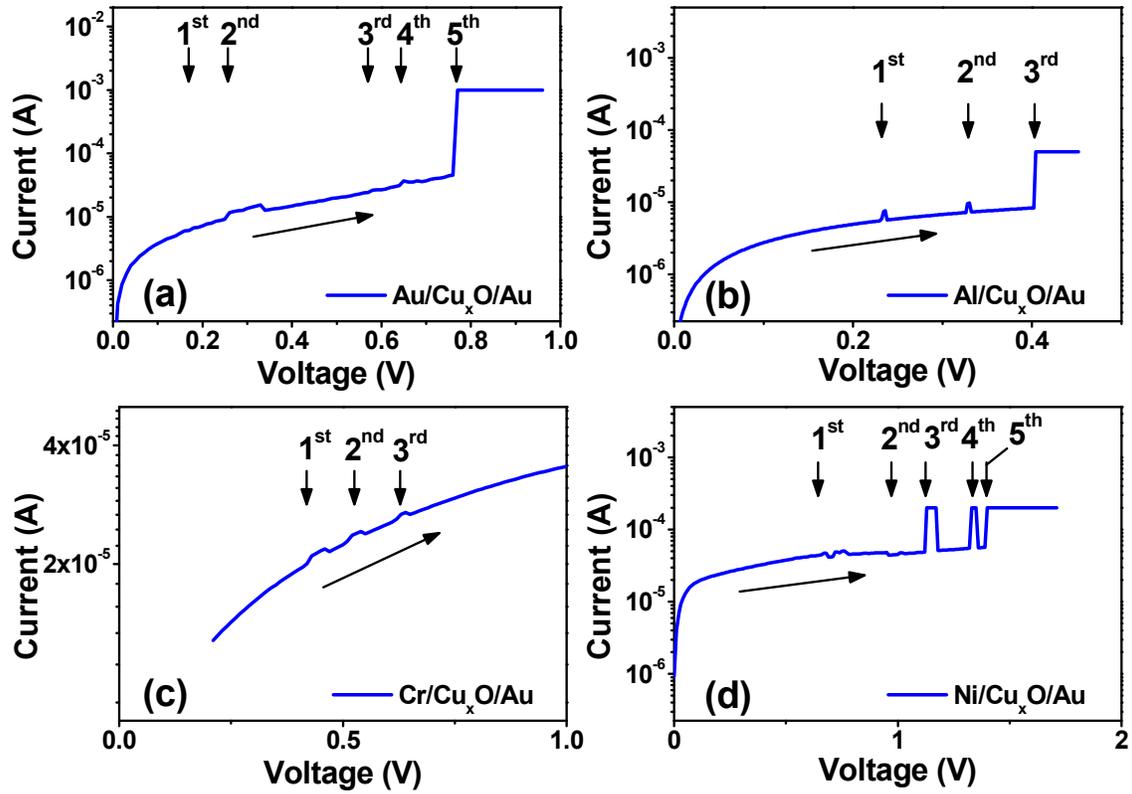

Figure 8 The number of pulses required for LIS to occur in (a) Au/Cu$_x$O/Au, (b) Al/Cu$_x$O/Au, (c) Cr/Cu$_x$O/Au and (d) Ni/Cu$_x$O/Au structures. The switch back from LRS to HRS in the intermediate pulses is to be noted.

Table 1 The total number of pulses incident on each device along with the number ($N_{LIS(1)}$) and percentage ($N_{LIS(1)}\%$) of devices that switched from the HRS-to-LRS in the first pulse

|  | Cr/Cu$_x$O/Au | Au/Cu$_x$O/Au | Al/Cu$_x$O/Au | Ni/Cu$_x$O/Au |
|---|---|---|---|---|
| LIS cycles | 97 | 15 | 27 | 9 |
| $N_{LIS(1)}$ | 82 | 14 | 18 | 3 |
| $N_{LIS(1)}\%$ | 84.5 | 93.3 | 66.7 | 33.3 |

A combination of inferences drawn from the results presented earlier are used to present a qualitative explanation for the mechanism of light induced resistive switching. It is known that insulator metal transitions, in Mott insulators such as CuO, are caused when the Fermi level $E_F$ moves above the mobility edge $E_c$ leading to the electronic states at $E_F$ becoming extended, which results in metallic conductivity [15]. This requires a critical charge density to be created by the incident light pulse. Thus, when the photogenerated carrier density reaches a critical value, the device makes a transition from HRS to LRS. Another mechanism that has been mentioned in literature is the drift of oxygen vacancies. In a previous work by the current authors, the origin of conduction in $Cu_xO$ has been established as a combination of Schottky emission and Poole-Frenkel defects [11]. Since $Cu_xO$ in the present case has a large number of defects, it is possible that under the influence of light illumination there is thinning of the barrier width for the transport of oxygen vacancies leading to the transition from HRS to LRS at low bias voltages In earlier work light controlled switching has been attributed to the mobility of light-modulated trapped electrons in the Schottky-like depletion layer [16,17].

In summary, the origin of light induced resistive switching appears to be due to photogenerated carriers that bridge the mobility edge under the influence of light illumination. The irreversibility of the process suggests that much higher intensity of light is required to reverse the process. The observation of reversibility in the case of Al top electrode would suggest that oxygen affinity of the top electrode plays an important role. The main advantage of this phenomenon is that it can be achieved using a low power white light source and even at extremely low forward biases of the order of tens of mV applied to the device. This is expected to open up a new range of applications including sensors.

**Summary**

Light induced resistive switching in M1/CuxO/M2 structures where M1 is Al, Cr or Ni and M2 is Au or Pt, is reported. Low power irradiation causes switching from HRS to LRS at very low voltages. Since the white light source is a normal halogen lamp or a low power LED source, the practical implementation is very simple. Significantly, in some cases the light pulse is introduced at ~0.4V which is much lower than the 2.5V for the set process in the absence of light. Furthermore the transition from HRS to LRS is irreversible suggesting the possibility of new type of applications. A qualitative explanation for the origin of light induced resistive switching is provided.

## Acknowledgments

The authors acknowledge facilities provided under the DST-PURSE programme and Central facility for nanotechnology of the University of Hyderabad.